\documentclass[article]{vgtc}  

\graphicspath{{figures/}{pictures/}{images/}{./}} 

\usepackage{times}                     

\usepackage{tabu}                      
\usepackage{booktabs}                  
\usepackage{lipsum}                    
\usepackage{mwe}                       

\usepackage{mathptmx}                  

\usepackage[normalem]{ulem}
\usepackage{verbatim}
\usepackage{amsmath}
\usepackage{caption}
\usepackage{hyperref}

\usepackage{xcolor}

\definecolor{mygreen}{RGB}{2,170,0}

\onlineid{1181}

\vgtccategory{Research}

\title{Perceptually-Guided Adjusted Teleporting: Perceptual Thresholds for Teleport Displacements in Virtual Environments}

\author{Rose Connolly\thanks{e-mail: connolr3@tcd.ie}\\ %
        \scriptsize Trinity College Dublin %
\and Victor Zordan\thanks{e-mail: vbzordan@roblox.com}\\ %
     \parbox{1.4in}{\scriptsize \centering Roblox Corporation \\ Clemson University}
\and  Rachel McDonnell \thanks{e-mail: ramcdonn@tcd.ie}\\ %
     \scriptsize Trinity College Dublin %
}

\teaser{
  \centering
  \includegraphics[width=\linewidth]{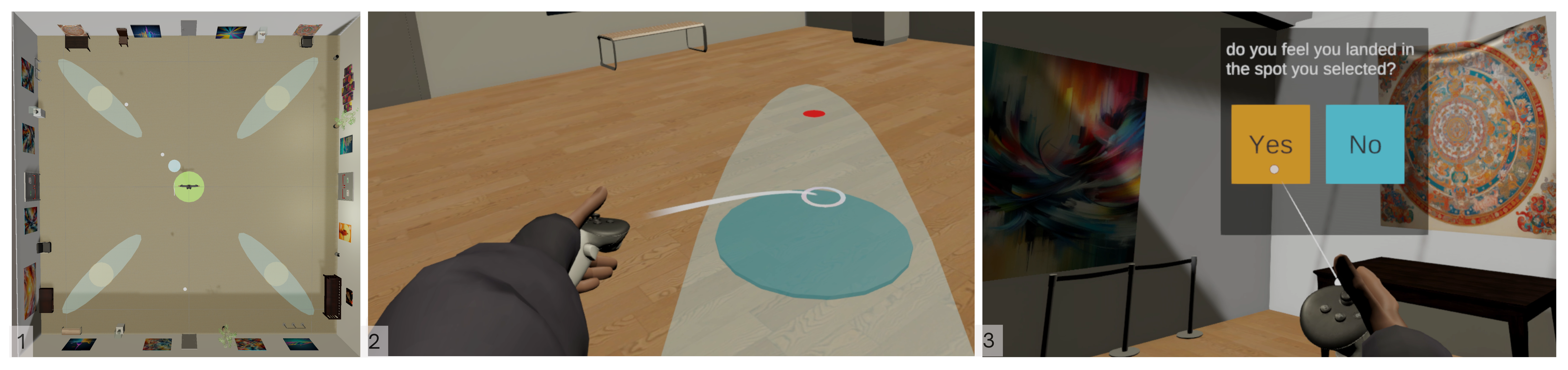}
  \caption{ 1. Bird's eye view of scene with teleport zone and possible destinations after adjustment visible. 2. Teleportation with adjustment of 1.1m indicated by red marker. 3. Question asked to user post teleportation. Please note adjustment area, possible teleport destinations and red marker are visible for illustrative purposes only, not visible during experiment.}
  \label{fig:teaser}
}

\abstract{
Teleportation is one of the most common locomotion techniques in virtual reality, yet its perceptual properties remain underexplored. While redirected walking research has shown that users’ movements can be subtly manipulated without detection, similar imperceptible adjustments for teleportation have not been systematically investigated. This study examines the thresholds at which teleportation displacements become noticeable to users. We conducted a repeated-measures experiment in which participants’ selected teleport destinations were altered in both direction (forwards, backwards) and at different ranges (small, large). Detection thresholds for these positional adjustments were estimated using a psychophysical staircase method with a two-alternative forced choice (2AFC) task. 

Results show that teleport destinations can be shifted without detection, with larger tolerances for backward adjustments and across longer teleport ranges. These findings establish baseline perceptual limits for redirected teleportation and highlight its potential as a design technique. Applications include supporting interpersonal distance management in social VR, guiding players toward objectives in games, and assisting novice users with navigation. By identifying the limits of imperceptible teleportation adjustments, this work extends redirection principles beyond walking to teleportation and opens new opportunities for adaptive and socially aware VR locomotion systems.

} 

\keywords{Virtual Reality, Locomotion, Teleportation, Perception.}

\begin{document}


\firstsection{Introduction}

\maketitle

Teleportation is one of the most widely used locomotion techniques in virtual reality (VR), advantageous in allowing users to traverse large environments without physical movement, overcoming space constraints \cite{boletsis2017new}. However, this comes with spatial navigation costs \cite{cherep2020spatial}. By bypassing natural sensory cues such as vestibular and proprioceptive feedback, teleportation disrupts spatial navigation \cite{campos2012multisensory, campos2014contributions}, impairs distance estimation \cite{keil2021effects}, and may contribute to disorientation. In social contexts, teleportation can also lead to unintended violations of personal space, causing discomfort when users appear too close to others \cite{connolly2025impact, wang2025can}. Thus, while teleportation solves one set of practical problems, it introduces new perceptual and social challenges.

Research on redirected walking has demonstrated that human perception in VR can be subtly manipulated without users noticing, enabling techniques that steer users safely within limited tracking spaces \cite{razzaque2005redirected}. These approaches rely on psychophysical thresholds to determine how much displacement can occur before detection. In contrast, imperceptible adjustments for teleportation have not been systematically examined \cite{prithul2021teleportation}. Prior research has primarily focused on augmenting traditional teleportation through explicit, noticeable modifications, such as user interface overlays or portal previews \cite{wolf2021augmenting, wang2025can}. Within redirected locomotion research, teleportation has occasionally been explored, but typically in portal-based systems designed to supplement walking rather than as a standalone technique \cite{liu2018increasing}. Consequently, the question of whether teleport destinations can be subtly altered (and to what extent) remains unanswered. Given teleportation’s inherent discontinuity and spatial consequences \cite{cherep2020spatial, keil2021effects}, it is an open question whether these characteristics could make users more tolerant of subtle adjustments than in continuous locomotion.

To address this gap, we conducted an experiment to measure the thresholds at which teleportation displacements become perceptible. Participants experienced systematic alterations to their teleport destinations, varying in both magnitude and direction (forwards, backwards). We employed a psychophysical staircase method, commonly used in redirected walking research \cite{congdon2019sensitivity, steinicke2009real}, to converge on detection thresholds. A two-alternative forced choice (2AFC) paradigm was used to ensure reliable measurement across two teleportation ranges (small and large). Our results show that teleport destinations can be shifted without detection, with larger tolerances for adjustments made backwards and across longer-range teleports. 

Prior research has demonstrated that guidance during teleportation can yield measurable benefits, such as increasing users’ likelihood of discovering hidden or overlooked objects compared to standard teleportation \cite{8864599}. However, these approaches rely on overt guidance mechanisms. In contrast, our findings establish baseline perceptual limits for imperceptible displacement in teleportation, extending the principles of redirection to a locomotion technique used by the vast majority of VR applications \cite{boletsis2017new}. As such, our study has direct implications for the design of teleportation in virtual environments.

Our proposed method of \textit{Perceptually-Guided Adjusted Teleportation} has potential to support personal space management, where users approaching a social partner via teleporting can be redirected to a comfortable distance from the partner (e.g., to a different zone of personal space \cite{hall1966hidden}). Given that users often teleport too close to others during interactions \cite{connolly2025impact}, subtle backward adjustments could help mitigate this tendency. These techniques may also benefit VR games by introducing subtle steering based on difficulty level e.g. guiding players toward objectives or assisting navigation in challenging areas, particularly for novice VR users. Thus, the distance and orientation inaccuracies commonly associated with teleportation \cite{keil2021effects, cherep2020spatial} could be mitigated through perceptually guided redirection. Such redirection could be applied more strongly for novice users and gradually reduced as users gain experience and spatial proficiency over time \cite{nasiri2023changes}, allowing assistance to adapt dynamically to player skill level.



By identifying the limits of undetectable teleportation adjustments, this study highlights the potential for more adaptive, comfortable, and socially aware teleportation systems. Our contributions are:

\begin{itemize}
    \item We introduce a perceptually-guided adjusted teleportation technique that subtly redirects teleportation endpoints while remaining imperceptible to users.
    \item We establish thresholds for adjusted teleporting in forward and backward directions, as well as at small and larger distances.
    \item We demonstrate how some of these thresholds vary depending on individual traits, providing guidelines for developers of VR locomotion systems.
\end{itemize}


\section{Background}
 Teleportation is one of the most common locomotion methods in VR, especially in commercial contexts~\cite{boletsis2017new}. Yet it remains under-researched compared to walking-in-place or continuous artificial locomotions \cite{boletsis2017new}. 

\subsection{Spatial Consequences of Teleportation}
While teleportation is effective at reducing motion sickness by avoiding continuous optical flow \cite{bowman1997travel}, it also introduces several negative spatial consequences in comparison to walking~\cite{cherep2020spatial}. As users do not visually traverse the space between their starting point and destination, they miss out on self-motion cues. This can result in spatial disorientation \cite{cherep2020spatial}. Additionally, teleportation makes it difficult for users to accurately estimate the distance travelled, often leading to underestimation \cite{keil2021effects}. As noted in prior work \cite{campos2012multisensory}, this discrepancy can occur in locomotion methods that lack proprioceptive and vestibular cues, which typically aid in distance estimation. Some of these drawbacks can be mitigated when teleportation is paired with real-world physical rotation, as opposed to controller-based rotation, since rotational visual cues remain intact \cite{kelly2020teleporting, cherep2020spatial}.

Additional adaptations to teleportation involve manipulating the visual transition that occurs during a teleport. Traditionally, teleportation is implemented as a discontinuous jump to the destination viewpoint. However, alternative transitions have been explored. Dash teleportation, for example, moves users rapidly through the space between the origin and destination, providing continuous optical flow \cite{bhandari2018teleportation}. This continuity supports path integration, which is disrupted by the discontinuity of standard teleportation \cite{cherep2020spatial,bhandari2018teleportation}. Importantly, dash teleportation preserves the primary benefit of conventional teleportation by not increasing VR sickness.

It has also been noted that there is little formal research that has classified teleportation into distinct ranges, such as small, medium, and large teleports \cite{yao2023does}. In most studies, teleportation distance is either left uncontrolled or emerges indirectly from the experimental setup rather than being treated as an independent variable. For instance, in one study \cite{kelly2020teleporting} the spatial effects of teleportation were examined by varying room size, which naturally caused teleportation length to scale with the environment \cite{kelly2020teleporting}. Although distance perception in VR has been studied in reviews \cite{renner2013perception, el2019distance}, its relationship with teleportation remains largely unexplored. In one of the few exceptions, teleport distances of 5m (small) and 15m (medium) were compared but no significant differences in spatial knowledge acquisition was reported \cite{yao2023does}. Another study defined short, medium, and long teleports as 7m, 14m, and 21m, respectively \cite{lee2023comparison}. This work reported decreased pointing accuracy with longer teleports, while efficiency, measured by aiming time, was longer for large teleports but showed no significant difference between short and medium ranges. Nevertheless, these findings leave open questions regarding smaller teleport ranges. This motivates research into perceptual qualities of finer-grained adjustments below 5–7m.

Research on distance estimation in VR provides additional context. Some studies suggest that estimation errors increase with distance \cite{gagnon2020far, ng2016depth}, while others \cite{saracini2020differences} found significant underestimation in peripersonal space (within arm’s reach), likely due to limited depth cues. Estimation errors in the spcae beyond one's immediate surroundings seem to be more pronounced in terms of underestimation \cite{armbruster2008depth,keil2021effects, ng2016depth}. Although findings are not entirely consistent, these results largely agree that estimation errors vary across distance ranges. Consequently, it is reasonable to expect that teleportation of different lengths may have distinct spatial effects.

\subsection{The Effects of Teleportation on Proximity}
Recently, the spatial consequences of teleportation have been examined in the context of proximity research \cite{wang2025can, connolly2025impact}, which studies how people use space socially \cite{hall1966hidden}. Teleportation appears to affect interpersonal distance (IPD) regulation. IPD, referring to the distance maintained between two people, is a key indicator of comfort in proximity research \cite{freeman2022disturbing}. A previous study found that users maintained shorter IPD when approaching agents via teleportation compared to walking, which they attribute to the difficulty of making small positional adjustments after a teleport \cite{connolly2025impact}. In everyday encounters, people can naturally adjust their positions to preserve comfortable spacing. However, in VR users may be more accepting of unnatural proximities, as making additional small teleports to refine spacing may not seem worthwhile. Yet research is needed into such fine-grained adjustments in teleportation and their associated cognitive costs.

As mentioned, recent research suggests that teleportation impairs IPD regulation \cite{connolly2025impact}. While the reasons for this are not explicitly clear, it would appear that when users teleport to another user from afar, it can be difficult to judge how close they will end up. The inaccuracies in this estimation can lead to an accidental overshoot, reducing IPD \cite{connolly2025impact}. A recent study has linked this disruption to personal space invasion anxiety \cite{wang2025can}. 

Social VR (SVR) in particular warrants attention, as intrusions into personal space can occur deliberately as a form of harassment. Such experiences may feel especially intense in SVR due to the heightened sense of presence \cite{blackwell2019harassment}. Although commercial SVR platforms have implemented strategies to manage personal space, these approaches have not been extensively studied in the literature \cite{mcveigh2019shaping}. For example, interviews with SVR designers revealed that most platforms include some form of personal space management: Rec Room, for instance, prevents users from teleporting too close to others. Other platforms use ``personal space bubbles” that render avatars invisible when they encroach on another user’s space. However, these features are often disabled by default, which can pose a challenge for inexperienced users who may be unaware of their existence or how to activate them \cite{mcveigh2019shaping}. It also remains unclear how effective these strategies are at supporting stable IPD regulation without compromising user comfort or agency in social VR.

Another approach to mitigating impaired IPD regulation is negotiated teleportation, as introduced in recent work \cite{wang2025can}. In this method, both parties can influence the final positioning before an interaction begins: one party selects an area the approaching user is “allowed in,” while the approaching user chooses a position within that area. This approach has the advantage that the resulting IPD reflects a form of aggregated preference between the two users. However, it introduces explicit UI interaction into the teleportation process, which may increase cognitive load. The interaction effort originates in the hands, rather than the legs, which has been shown to lead to arm fatigue \cite{lou2020empirical}. Yet a study did find that teleportation had a lower cognitive load compared to non-controller methods \cite{griffin2018evaluation}. However others argue that teleportation should increase cognitive load, especially since users in many VR environments also need to use controllers for other types of interactions \cite{laviola2001hands} \cite{griffin2018evaluation}. Therefore the addition of UI methods to a locomotion method that already has controller based input (unlike natural walking) may be cognitively taxing. Thus negotiated teleportation may reduce personal space invasion anxiety, but may impose additional burden. This motivates exploring imperceptible adjustments: subtle manipulations that support natural social spacing without requiring user effort.

\subsection{Redirected Locomotion Techniques}
A large body of VR research has explored redirected locomotion techniques, which exploit perceptual thresholds to manipulate either the user’s movements or the virtual environment without detection \cite{razzaque2005redirected, suma2012impossible,nilsson201815}. 

Perhaps the most well-known and extensively studied of redirected techniques is redirected walking (RDW) \cite{razzaque2005redirected}. RDW manipulates the user’s head-mounted display (HMD) by scaling head rotations or subtly curving walking trajectories, enabling traversal of large virtual spaces within limited physical tracking areas \cite{razzaque2005redirected}. Considerable research has since refined and extended this technique (see review by \cite{fan2022redirected}). For instance, it has been demonstrated that incorporating eye blinks into the redirection process can increase redirection effectiveness by up to 50\% \cite{langbehn2018blink}. Another study, targets manipulation during saccades (rapid eye movements). During a saccade of 15\textdegree, the scene could be rotated up to 5\textdegree and translated forwards or backwards by 0.5m \cite{bolte2015subliminal}.


%

Redirected locomotion techniques are commonly categorized into those that manipulate the user's own movements and those that manipulate aspects of the virtual scene \cite{suma2012impossible}. RDW as introduced in its original form \cite{razzaque2005redirected} involved manipulation of the mapping between the user’s real and virtual rotation \cite{nilsson201815}. However other approaches introduce manipulation of the virtual room \cite{nilsson201815}. For instance, Suma et al. \cite{suma2011leveraging} proposed repositioning virtual objects, such as doors, when users are distracted, exploiting change blindness to allow the reuse of physical areas across multiple virtual segments. Similarly, “impossible spaces” make use of non-Euclidean layouts to overlap or rotate parts of the environment \cite{suma2012impossible}. Although these layouts cannot exist in the physical world, they are readily accepted in VR and can be effectively leveraged for redirection \cite{suma2011leveraging, lutfallah2024perception}. As well as during eye blinks or saccades, discrete redirection has also been applied during user interactions, such as an object pickup \cite{schmelter2021interaction}. Thus discrete manipulations can leverage the change blindness effect for redirection purposes \cite{schmelter2021interaction, suma2011leveraging}.

In contrast, continuous redirection gradually modifies the user’s ongoing movements through gains applied to their walking, subtly steering users without detection \cite{coles2025exploring}. Many RDW approaches use combinations of various techniques \cite{liu2018increasing}. A comprehensive review of perceptual limits in redirected walking highlights a range of such continuous gain modalities, extending beyond early rotational gains to include translational and bending gains \cite{coles2025exploring}. Reported detection thresholds for translational gains vary across studies: Steinicke et al.\ identified imperceptible gain ratios of $0.78$–$1.22$ \cite{steinicke2008analyses}, later refined to $0.86$–$1.26$ \cite{steinicke2009estimation}, while Kim et al.\ demonstrated environment-dependent thresholds ranging from $0.73$–$1.10$ in small empty rooms to $0.91$–$1.22$ in large empty rooms \cite{kim2023effects}. Overall, these findings indicate no consistent asymmetry between increasing or decreasing walking distance, with imperceptible translational gains generally clustering around approximately $\pm20\%$ \cite{kim2023effects, steinicke2008analyses, steinicke2009real}.






\subsubsection{Determining Perceptual Thresholds with psychophysical methods}
Central to these approaches is the determination of perceptual thresholds, the limits of undetectable change a user can experience \cite{5072212, lutfallah2024perception, suma2012impossible}. Subtle manipulations of motion cues often go unnoticed, enabling efficient reuse of limited physical space \cite{congdon2019sensitivity}. Such redirection detection thresholds (RDTs) can be determined using a variety of psychophysical methods. 

The method of constant stimuli (MCS), for example, presents participants with a wide range of stimulus intensities to construct a psychometric function \cite{pelli1995psychophysical}. As the primary approach in the first comprehensive study on RDTs for redirected walking (RDW) \cite{steinicke2009real}, MCS remains a common choice in RDT experiments \cite{chen2019quick}. In contrast, the method of adjustment (MoA) allows participants to modify the stimulus themselves until it matches a given reference \cite{treutwein1995adaptive}. Chen \cite{chen2019comparison} reported that MoA yields results comparable to MCS while reducing experimental time by approximately 33\%.

 As the MCS is time-consuming, a variation of the method of limits is the staircase method, an adaptive method involving both ascending and descending limits in a trial \cite{cornsweet1962staircase}. As the stimulus intensity is adjusted based on participant responses, using both ascending and descending sequences, which avoids presenting stimuli far above or below the threshold \cite{cornsweet1962staircase, treutwein1995adaptive}. The staircase method has been successfully applied to measure thresholds in redirected walking tasks, such as accelerated physical rotations \cite{congdon2019sensitivity} and subtle alterations of walking direction \cite{steinicke2009real}.


Of course the disadvantage of such methods is the participant is aware of the experiments purpose. When users engaged in a task not related to manipulations they are more tolerant of redirection \cite{nguyen2020effect, steinicke2009estimation}. Furthermore, thresholds vary across individuals. Congdon et al. \cite{congdon2019sensitivity} showed that experienced VR users are more sensitive to redirection manipulations than novices. These findings highlight that both individual differences and situational context play a critical role in whether manipulations are detected.


\subsection{Redirected Teleportation}
However, nearly all redirected techniques focus on walking-based locomotion. Since their primary purpose is to bypass physical space constraints, teleportation is often neglected as by its nature it already bypasses these constraints by instantly relocating the user. Yet as an inherently spatially disorientating method (\cite{cherep2020spatial}) teleportation may have unique potential for redirection.

A review of teleportation techniques in VR (\cite{prithul2021teleportation}) highlights only one redirected teleportation~\cite{liu2018increasing}. This approach employs `portals' that subtly shift the user’s position as they cross a boundary, effectively augmenting natural walking rather than altering the traditional point-and-move teleportation mechanic. Another related approach is `directed teleportation', in which users are rotated during teleportation toward points of interest, without affecting their final position \cite{8864599}. This method is explicitly perceivable to the user, unlike subtle redirection techniques. Nonetheless, their results demonstrated benefits: participants were more likely to discover hidden or overlooked objects compared to standard teleportation. While informative, this work focuses on overt guidance rather than subtle manipulation.

In commercial contexts, similar concepts exist in the form of teleportation anchors, such as those provided by the OpenXR Toolkit\footnote{\url{https://docs.unity3d.com/Packages/com.unity.xr.interaction.toolkit@2.0/manual/teleportation-anchor.html}}, which move users to predetermined positions or orientations. However, these manipulations are explicit and visible to the user, unlike subtle redirections. Other explicit teleportation augmentations, such as teleport with rotation \cite{wolf2021augmenting}, similarly make the change obvious to the user rather than covertly influencing navigation or perception. As mentioned, personal space bubbles have been utilised in SVR, but again serve as a perceivable feature, where the user vanishes if you intrude upon them \cite{mcveigh2018s, mcveigh2019shaping}. 

Collectively, these observations suggest that while explicit modifications of teleportation have been widely implemented and studied, subtle manipulations of traditional teleportation remain a largely open area for investigation. 

Although teleportation has rarely been studied through the lens of redirection, it presents unique opportunities. The spatial disorientation it induces (\cite{cherep2020spatial}) may actually broaden perceptual thresholds, enabling manipulations that would be noticed in walking contexts. In other words, teleportation’s inherent disruption could be repurposed as a design advantage. At the same time, individual differences in spatial ability must be taken into account, as users vary considerably in how they navigate and reorient after teleportation \cite{cherep2023individual}.

Other considerations must be taken into account when considering manipulation of teleportations. For instance the landmarks present in the scene, whose presence reduces disorientation and improves wayfinding in teleportation contexts \cite{sharma2017influence}. Proximal (close) landmarks provide positional information, while distal (far) landmarks can offer directional guidance \cite{padilla2017sex}. Boundaries, including walls and fences, or even flooring changes, act as spatial anchors that reduce disorientation after teleportation \cite{kelly2021boundaries, cherep2020spatial}. Enriched spatial contexts similarly aid spatial updating and orientation \cite{borodaeva2023spatial}. Thus, while teleportation disrupts spatial updating, carefully designed environmental cues can partially offset these deficits \cite{kelly2021boundaries, borodaeva2023spatial,sharma2017influence}.

While extensive research has established perceptual limits for RDW \cite{fan2022redirected, coles2025exploring}, it remains unclear if and how these limits can transfer to redirection of teleportation-based locomotion. The discrete nature of teleportation may render commonly reported continuous gain ratios and perpetual thresholds incomparable. Discrete forms of redirection may be more applicable and have been explored in some aspects of RDW to enable brief, imperceptible spatial adjustments \cite{schmelter2021interaction, suma2011leveraging, langbehn2018blink}. However, teleportation presents a unique challenge in that it already consists of a discrete, expected redirection (the teleportation itself). A teleportation undergoing adjustment would combine this with an additional imperceivable redirection. As such, comparing perceptual thresholds for RDW to teleportation-based adjustments may be challenging, as they differ in both continuity and user expectation.


Overall, redirection of traditional point-and-move teleportation remains unexplored but offers distinct opportunities for user guidance in VR, yet its success may depend on environmental design and individual spatial skills.




\section{Objectives \& Hypotheses}

Teleportation is  a locomotion method in which the user may specify a destination in the virtual environment by pointing. This is usually done using a handheld motion tracked controller. When activated, the user’s viewpoint may instantly transition to the specified destination \cite{laviola20173d}.

Thus we define \textit{adjusted teleportation} as a redirected locomotion method in which the user's specified destination is altered in some way. This is not to be confused with \textit{redirected teleportation}, \cite{liu2018increasing} which imperceptibly redirects the user’s virtual position during natural walking. The scope of this this paper is limited to forwards and backwards translational adjustments, but we acknowledge there is potential for other directional, or even rotational adjustments.

Our primary objective was to determine whether teleport destinations can be manipulated in the forwards or backwards direction, and if so, by what magnitude. We also sought to examine whether such thresholds, if present, differ depending on teleportation length. Previous work has shown thresholds for translation thresholds to be 0.5m in cases without movement  \cite{bolte2015subliminal} (during a saccade). We used this as a reference because it provides a discrete, non-continuous comparison, in contrast to gradual translation manipulations. With the addition of locomotion, we predicted that this threshold would increase, as the user already expects some degree of translation.

In addition to the above objectives, we also devised a set of hypotheses guided by prior work:
%
\begin{itemize}
    \item \textbf{H1:} Perceptual thresholds for adjusted teleportation will be larger for backwards adjustments than forward adjustments. 
\end{itemize}
When teleporting, users have an expected post-teleport viewpoint. Environmental objects act as piloting cues for these viewpoints, making forward adjustments more noticeable as participants may move closer to or past these objects. In contrast, backward adjustments will be less likely to be detected because the objects remain in view, keeping the expected viewpoint largely intact.


\begin{itemize}

     \item \textbf{H2:} Perceptual thresholds for adjusted teleportation will be larger when the teleport range is large, compared to small
\end{itemize}

Prior work suggests decreased pointing accuracy with large teleports \cite{lee2023comparison}, as well as decreased distance estimation abilities with larger distances \cite{gagnon2020far, ng2016depth}. As such, we predict users will have more difficulty predicting their position following a teleport when it is further away.


\begin{itemize}
           \item \textbf{H3:} Users who are more experienced in VR will have smaller perceptual thresholds for adjusted teleportation.
\end{itemize}

              A study by Congdon et al. (2019) \cite{congdon2019sensitivity} showed that experienced VR users are more sensitive to  manipulations in redirected walking than novices. We predict the same pattern in our results. 
\begin{itemize}
      \item \textbf{H4:} Users with higher spatial abilities will have smaller perceptual thresholds for adjusted teleportation.
\end{itemize}
   
Prior works has shown correlation of spatial ability with performance in other teleportation tasks \cite{cherep2023individual}. As such we predict a similar trend for our study. 

\section{Experimental Setup}
Participants used the Oculus Quest 3 Head-Mounted Display (HMD)
for the experiment, which was developed in Unity 6000.0.047f1. Participants embodied an avatar from Ready Player Me\footnote{https://readyplayer.me/}, designed with a gender-neutral appearance with baggy clothing to hide the body shape. The floor was used as the tracking origin and the  virtual body’s height matched the participant’s height.
\subsection{Adjusted Teleports}
 For each trial, participants were instructed to teleport within a given range. The available teleportation area was a circle of radius 1m, differentiable from the rest of the floor by a semi-transparent blue hue. The first 10 teleports were used as practice teleports and were disregarded to discount for any spatial learning effects. No adjustment occurred in the training trials.

Teleportation was implemented with a curved trajectory ray and a reticle that indicated the selected destination. Teleportation was triggered with a controller input. For adjusted teleports, the destination was instantly altered from the selected one by both magnitude and direction (forwards or backwards) when the user triggered the teleport.  

The adjustment direction was determined relative to the player's intended movement, from their current position (origin) to their selected destination. Using the vector from the player’s origin to their destination as the forward reference, directional adjustments took place either forwards or backwards. Adjustment magnitude was calculated from the user's selected destination.  


\subsection{Teleport Ranges}
We designed a repeated-measures study involving small and large teleports. Prior work provides limited classification of teleport distances into ranges \cite{yao2023does}. Small teleports have been defined as 5 m \cite{yao2023does} or 7 m \cite{lee2023comparison}, but as our study aimed to investigate smaller-scale adjustments we chose a range of 2.5m as small. The larger range was chosen to be noticeably greater than this baseline, providing a clear perceptible difference from the small teleports while still remaining within a realistic VR movement scale. Based on these considerations, we defined our teleport ranges as:  \textit{Small:} 2.5m, and \textit{Large:} 9m.

\subsection{Perceptual Thresholds with the Staircase Method}

We used the Unity Staircase Procedure Toolkit  \cite{10.1145/3611659.3617218} to determine the Point of Subjective Equality for forward and backwards adjustments across both small and large teleport ranges. Instead of method of constant stimuli paradigm which would have resulted in many trials and a coarse threshold, we selected staircase procedure to hone in on the threshold quicker \cite{cornsweet1962staircase, treutwein1995adaptive}. A staircase procedure adjusts stimulus intensity based on a participant’s responses to determine their detection threshold. 
The procedure works using two inner staircases, with one starting from a strong stimulus (an obvious, detectable and large adjustment) and one from a weak stimulus (an undetectable small adjustment).

After each teleport, a UI appeared which asked the the following question
\textit{``Do you feel you landed in the spot you selected? (yes/no response)"}. This UI was both semi-transparent  and appeared 1.5 seconds after their teleport, in order to allow them time to view their surroundings without the UI. Participants were instructed not to turn around before responding. Although there was no time limit, they were asked to provide their answers based on their initial, intuitive judgment. After each teleport, the stimulus becomes stronger if it was not detected and weaker if it was detected, with reversals marking changes in response. 

\begin{figure}
    \centering
    \includegraphics[width=0.99\linewidth]{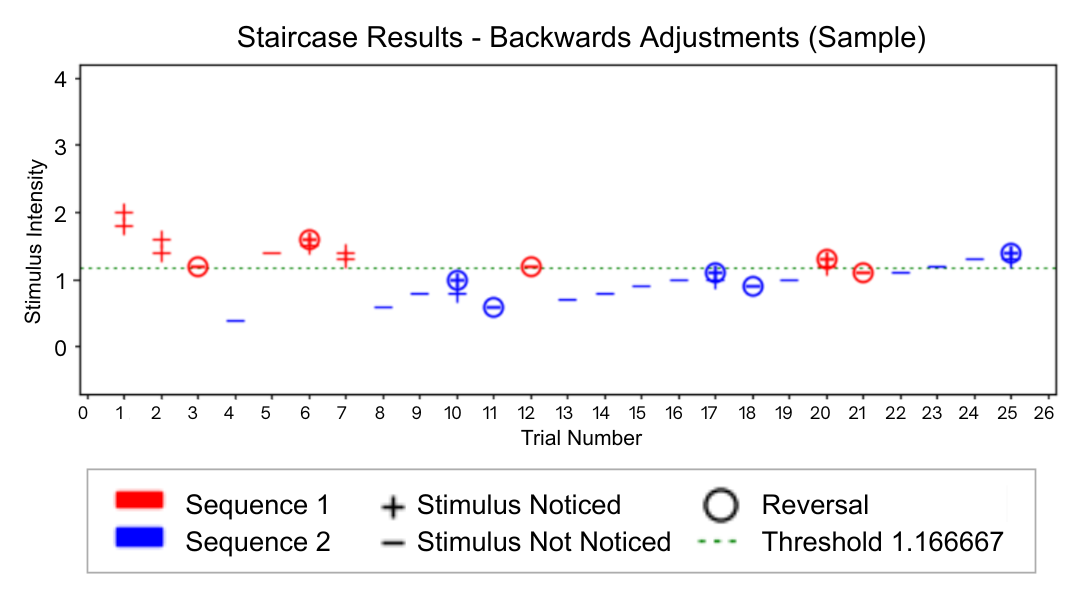}
    \caption{Sample staircase procedure showing convergence at a threshold of 1.67. 
}
    \label{fig:sample}
\end{figure}

Each staircase continued until convergence at a final threshold. The Unity Staircase Procedure Toolkit parameters were adjusted to determine the threshold when 5 reversals had occurred in both the upper and lower staircases, as per \cite{feick2021visuo}. Please see figure \ref{fig:sample} for an example staircase. The final threshold is then the averaged value of last three of the reversal stimuli's values \cite{garcia1998forced}. 

The Unity Staircase Procedure Toolkit also offers additional parameters to refine the staircase method. Based on informal pilot testing, the starting values were set to 2m for the upper staircase and 0.8m for the lower staircase, which facilitated faster convergence \cite{treutwein1995adaptive}. To target the 50\% point on the psychometric function, a 1-up-1-down staircase is sufficient \cite{10.1145/3611659.3617218}. Therefore we employed an 1-up-1-down staircase procedure, designed to converge at the stimulus level corresponding to a 50\% probability of correct detection. In this method, the stimulus intensity is increased following incorrect responses and decreased following correct responses, using the same step size. Additionally, a “quick start” was implemented, in which the step size was initially doubled until the first reversal occurred, accelerating convergence.

For each block participants underwent two separate staircase procedures for both forward and backward adjustments. The staircase was chosen randomly until each staircase concluded. The adjustment magnitude could potentially range from 0m (no adjustment) across both teleport ranges to 2.5m for small range teleports and 3.5m for large range teleports. The maximum adjustment for large teleports was chosen based on informal pilot testing to determine a universally noticeable adjustment. For small teleports, the maximum adjustment corresponded to the largest possible change without moving participants backward from their starting position.

During each block, a control condition was introduced in which no adjustment occurred. There were 10 of these trials introduced, serving as catch trials to rule out inattentive participants. Thus in each block, trials alternated randomly between three conditions: forward adjustments, backward adjustments, and no adjustment. The block concluded when both staircases had converged, and all control trials had been completed.

\subsection{Virtual Environment}
The experimental environment was a square virtual room. 
Users completed two blocks (small and large teleport ranges) in counterbalanced order. For each teleport range users were instructed to teleport within a defined circle of radius 0.5m, allowing them some flexibility over the teleport, but still constraining them to a certain range. There were 4 possible places the teleportation zone could be, arranged in a square formation with the user start position in the centre. Please see figure \ref{fig:teaser} (1), for layout of room (large condition). The teleport zone was positioned randomly to one of these 4 possible areas for each trial, each area at a constant distance from the centre of the room. The centre of the teleport ranges, and the possible range of teleports are given below:
\begin{itemize}
    \item Small: 2.5m (allowing range of 2-3m)
    \item Large: 9m (allowing range of 8.5-9.5m)
\end{itemize}

In each block the boundary wall of the environment were 4.5m away from the nearest edge of the teleport zone. This was to allow ample room in case of large adjustments. This meant the room for each block had different lengths (approximately 13.54m and 22.72m respectively), with the ceiling size constant (6m).  The teleport zone was randomised for each trial and the participant returned to the centre (a circular area also of 0.5m radius) via teleportation after each teleport. The teleport zone was no longer visible after the player teleported as it would serve as an artificial positioning cue.

\subsubsection{Piloting Cues}
\begin{figure}
    \centering
    \includegraphics[width=0.99\linewidth]{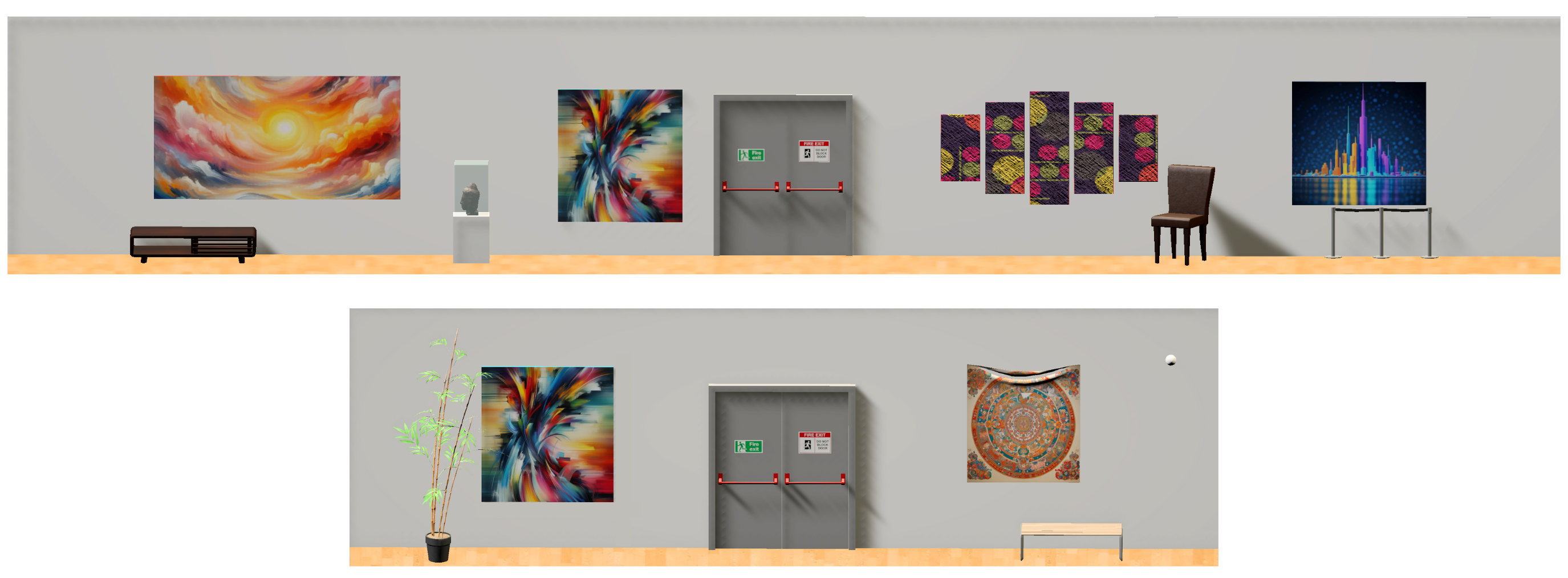}
    \caption{Layout of wall for rooms in large teleport condition (above) and small teleport condition (below)}
    \label{fig:walls}
\end{figure}

The layout of the virtual room was inspired by \cite{kelly2020teleporting}, who designed both large and small versions of a warehouse-style environment, with objects arranged along the walls and the center space left open. We adopted a similar layout in an art gallery style. In our design, objects such as plants, chairs, and art displays were placed symmetrically along the walls to serve as piloting cues, while leaving the central area unobstructed. Each wall also featured a centrally located door, accompanied by two or four paintings depending on room size. All items were taken from the ``Museum Interior"\footnote{https://assetstore.unity.com/packages/3d/props/interior/museum-interior-281559} asset on Unity Asset Store. Symmetry was maintained across both axes, so that each corner of the room contained similarly positioned cues, though their types varied to add visual interest and realism. In the small-room condition, each wall displayed two paintings and two objects, whereas in the large-room condition, each wall displayed four paintings and four objects. An example wall and object layout is shown in Figure \ref{fig:walls}.


\section{Experimental Design}

\subsection{Participants}
Prior to our experiment, we conducted an a priori power analysis
using G*Power 3.1.9.7, which is a widely used statistical software program
designed for sample size calculation in experimental and behavioural
research, to determine the appropriate sample size for our study. In order to detect a medium effect size of 0.25, with power $(1 - \beta)$ of 0.8 and $\alpha$ of 0.05, G*Power suggested that we would need
24 participants for a repeated measures ANOVA.

We recruited 32 participants (19 male, 13 female) for our study, each receiving a book voucher worth €10 as compensation. The study was approved by our university's research ethics committee.
Our participants were aged 23-61 ($M : 29.61, SD : 9.35$).

\subsection{Procedure}
\label{sec:Procedure}
At the beginning of the experiment, each participant took a questionnaire that that measured their spatial and navigational ability. We used the Santa Barbara Sense of Direction Scale (SBSOD) \cite{hegarty2002development}. This involves 15 questions such as \textit{`I am very good at giving directions'} or \textit{`I can usually remember a new route after I have travelled it only once'} which are scored to provide a value where 1 indicates a poor sense of direction and 7 indicates a good sense of direction. We also asked for demographic information including their gender and experience with VR. VR experience was rated on a scale from 1 to 5, with 1 indicating a novice user and 5 indicating an expert.

We then informed the participants of the procedure and showed them the controller inputs they would need to use. Then, they were asked to put on the HMD, adjust until comfortable and to hold the controllers. Participants underwent two blocks in which the teleport range differed (small or large). After the first block they rested for 5 minutes, or longer if desired. Each block began with ten training trials (no adjustment) to ensure participants were familiar with the controls and to allow any spatial learning benefits from navigating the environment to occur before data collection began. Data from these training trials were not included in the staircase procedure.

We also included catch trials (no adjustment) throughout the experiment to determine if any participant has an excessive tendency to detect adjustments when there was none, or was inattentive in their response. Participants scoring below 70\% on catch trials were excluded, as per with prior work using similar performance-based cutoffs \cite{matzen2024machine, son2022scene}.

For each block, the participants began in the center of the virtual room. They then had to teleport anywhere in a teleportation range: visible as circular area (radius of 1m) with a transparent blue glow. The teleport range randomly alternated from each corner of the room, to maintain consistent boundary walls. During each block 2 staircases were running, which adjusted the teleport forwards, or backwards accordingly. Each trial randomly alternated to a different staircase until each was completed. After each trial's teleport participants answered ``Do you feel you landed in the spot you selected?" via a virtual UI menu.  The teleportation range was not visible after the teleportation took place, to prevent the use of the teleport range as a boundary cue. Participants could freely turn their heads, while controller-based rotation was provided for body orientation. This approach ensured a smooth and consistent experience, as participants were expected to teleport in straight lines without extensive physical reorientation.

Participants teleported back to the same start location after each teleport, whereupon the next teleport range became visible. This teleport back to the start location was a `hotspot' teleport meaning that the participants could only teleport to a single position, as such it differed from the teleport `ranges' which offered more freedom to teleport anywhere within a given range. The participants could not teleport elsewhere in the room: they could only teleport to the available teleport range during a trial and only teleport back to the start after completing a trial. 

After finishing both virtual reality blocks, participants filled out a questionnaire where they were asked via an open text box if they used any particular technique to make their decision. They then underwent a computerised version of the spatial orientation test (SOT) \cite{friedman2020computerized} to determine their perspective-taking and direction-estimation ability. The SOT yields a performance measure based on the participant’s average absolute angular error, with lower errors indicating better perspective-taking ability.

We included this test at the end of the study because informal pilot testing suggested that completing it before the teleportation tasks could help participants develop strategies for noticing adjustments.

\section{Results}
1 participant was ruled out for failing the catch trials  (less than 70\%), leaving 31 participants total.

\begin{figure}[t]  
    \centering
    \includegraphics[width=0.5\textwidth]{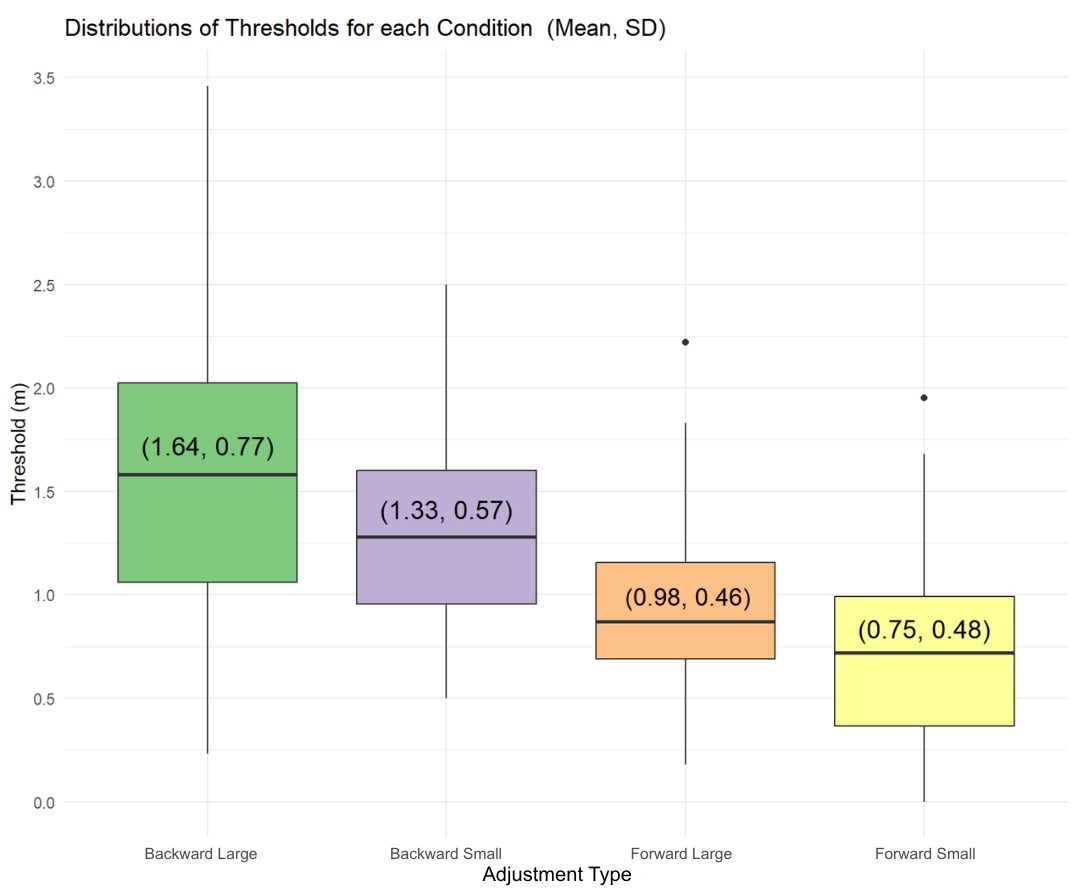}
    \caption{Distributions of thresholds for each condition with mean and standard deviation labelled (measured in meters), ordered largest to smallest.}
    \label{fig:distr}
\end{figure}

The average threshold value for backward adjustments was 1.64m across the large range condition, and 1.33m in the small. The thresholds were 0.98m and 0.75m for forward adjustments during large and small conditions, respectively. See figure \ref{fig:distr} for boxplot of each condition.

\subsection{Difference in Size and Direction Thresholds}
We used a two-way repeated measures ANOVA to examine differences in thresholds across adjustment direction (forwards vs. backwards) and teleport size (small vs. large). The data had no outliers (values outside the range $[Q1 - 3 × IQR, Q3 + 3 × IQR]$ were classed as outliers). Normality was checked for each condition using the Shapiro-Wilk test. As the normality assumption was violated in the forwards-large condition, we applied an Aligned Rank Transform to the data using the R package ARTool \cite{10.1145/1978942.1978963} (as used in studies: \cite{electronics12143051, vargas:hal-04369761}).  Statistical significance was evaluated at $p < 0.05$.

There was a significant main effect of \textit{direction}, $F(1,90)=46.11$, $p<0.001$, $\eta^2_\text{partial}=0.34$, and a significant main effect of \textit{size}, $F(1,90)=7.43$, $p=0.008$, $\eta^2_\text{partial}=0.08$. The interaction was not significant.








\subsection{Correlation with Personal Measures Spatial Ability}

We conducted Pearson’s correlation between each condition (forward and backwards adjustments, small or large teleport range), their spatial ability scores (SOT and SBSOD) and their experience level with VR.

We adjusted for multiple comparisons using the False Discovery Rate. Their was significant positive correlation between SOT and backwards adjudgements across large teleport ranges (p=0.03). There was significant negative correlation between VR experience level and backwards adjudgements across small teleport ranges (p=0.03). 

See Table \ref{tab:correlations} for Pearson’s correlations of each condition with SOT and SBSOD scores, as well as experience with VR. Higher SOT scores indicate poorer perspective-taking, higher SBSOD scores indicate a better sense of direction, and VR experience was coded from 1 (least) to 5 (most).

\begin{table}[ht]
\centering
\small  
\setlength{\tabcolsep}{4pt} 
\caption{Correlation coefficients, significant correlations are marked with *. }
\begin{tabular}{lcccc}
\hline
 & Forward (Small) & Forward (Large) & Back (Small) & Back (Large) \\
\hline
SOT   &  0.11 & -0.13 &  0.38 &  0.49* \\
SBSOD & -0.35 & -0.13 & -0.21 & -0.10 \\
VR Experience   &  0.03 &  0.18 & -0.49* & -0.30 \\
\hline
\end{tabular}
\label{tab:correlations}
\end{table}

\subsection{Comments}
We also asked particpants whether they used any particular technique in making their decision. 27 provided answers. Of these, 19 mentioned using the objects or walls as landmark references. 7 specifically mentioned aspects of distance estimation, mostly to objects or walls. 7 mentioned aspects of triangulation, examining their relative alignment to objects. Examples are given below.
\begin{quote}
   \textit{ ``I tried to visualise how the different objects would be in relation to the new location that I was teleporting to."}
\end{quote}

\begin{quote}
     \textit{  ``Predicting how things would look just before releasing the button, then comparing my surroundings with my mental image."}
\end{quote}
\begin{quote}
   \textit{``Using pictures/objects at the wall as reference points. if i aimed before a poster and landed after it was jarring and i usually answered no to the question."}
\end{quote}
\begin{quote}
      \textit{ ``I looked ahead at the objects to try to understand at what distance I should land from the table, for instance."}
\end{quote}

\section{Discussion}

The aim of this study was to determine whether teleportation destinations can be manipulated in either the forward or backward direction without participants noticing the displacement, and by how much. Using a staircase procedure, we were able to measure adjustment thresholds for both directions, across two teleport ranges.

We had expected thresholds for teleport adjustments to at least measure 0.5m, given a study that measured this as the threshold
for discrete translation thresholds without movement \cite{bolte2015subliminal} (during a saccade). However with the addition of locomotion, we predicted that this threshold would significantly increase, as the user already expects some degree of translation. As such, even the minimum threshold we measured, 0.75m (for forwards adjustments across small teleport ranges) is a significant distance offering potential for adjusted teleportation. This indicates that when users anticipate a translation i.e. during a teleportation event, this expectation can be exploited to lengthen, or shorten, their teleportation. Additionally, teleportation is widely accepted as having negative consequences for spatial orientation and general distance estimation \cite{cherep2020spatial,keil2021effects}. This spatial disorientation may contribute to the perceptual limits observed, potentially allowing greater flexibility in adjusting teleportation without detection.



While our primary objective was to explore whether adjusted teleportation was feasible as a concept in itself, we also had a number of hypotheses in mind. With our first hypothesis (\textbf{H1}), we predicted that teleport destinations would have larger perceptual thresholds for backward adjustments compared to forward adjustments. Our results supported this hypothesis: backward adjustments yielded larger thresholds across both small and large teleport ranges.





\begin{figure}
    \centering
    \includegraphics[width=0.99\linewidth]{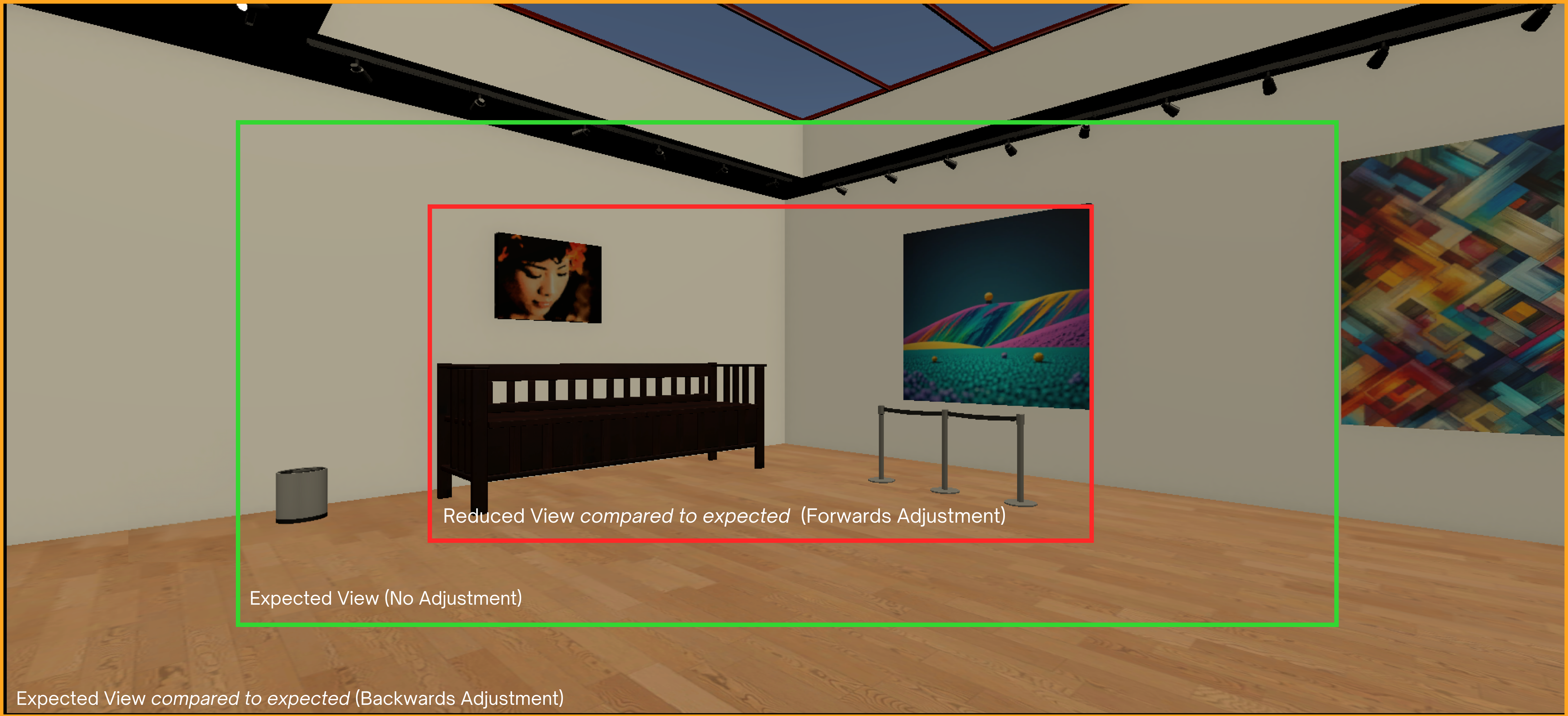}
    \caption{Captured camera positions for reference, showing: (1) backward adjustment of 1.5m (outer/orange outline), (2) no adjustment (middle/green outline), and (3) forward adjustment of 1.5m (inner/red outline).}
    \label{fig:fov}
\end{figure}

Some participants reported anticipating their viewpoint prior to teleportation in order to detect adjustments. Differences in sensitivity between forward and backward adjustments may therefore be explained by the expected versus actual field of view (FOV) following an adjusted teleport. When a teleportation destination is adjusted backward, much of the originally expected scene remains within the FOV, preserving visual continuity. In contrast, a forward adjustment effectively ``cuts off” portions of the expected FOV, introducing novel elements that were not anticipated (see Figure \ref{fig:fov}). This lack of overlap may make forward displacements more noticeable and easier to detect. However, as our findings are limited to an enclosed space, it remains unclear whether this effect would generalize to open-world environments.

Additionally, many participants reported using objects as positional cues to estimate their post-teleport location. Following a significant forward adjustment, these reference objects may be partially or completely absent from the FOV, differing markedly from expectations. By contrast, backward adjustments tend to preserve the presence of these objects within the FOV. While we did not compute the pixel overlap of the headset’s view or weight the frame of view at each adjustment point, future research could investigate whether these factors contribute to perceptual differences between forward and backward teleport adjustments. Please see Figure \ref{fig:fov} for a comparison of camera positions after an unadjusted teleport, as well as both a forwards and backwards adjusted teleportation.

Previous studies also suggest that teleportation can sometimes result in accidental overshoot \cite{connolly2025impact}, which can be particularly uncomfortable if it occurs near another user, potentially triggering personal space invasion anxiety \cite{wang2025can}. In this context, forward adjustments could increase discomfort, whereas subtly moving a user backward may even be subconsciously welcomed. Considering that backwards adjustments have larger thresholds, teleport destinations could even be used to prevent users from interrupting another's personal space, especially when they tend to overshoot, so using backwards adjustment as a correction would be feasible. For backwards teleports, even for small teleport ranges, a distance of 1.33m could be sufficient to correct a teleportation that invades a person's intimate space (typically ($<0.45m$ \cite{hall1966hidden}) back into more appropriate personal spaces zones, i.e. the personal , or even social zone (defined as $0.45m-1.2m$,  $1.2m-3.6m$ respectively \cite{hall1966hidden}).

We also found differences across teleport ranges. Teleport destinations could be adjusted further when the teleport range was large, compared to small, supporting \textbf{H2}. A previous study found that pointing accuracy was highest for short teleports, followed by medium and then large distances \cite{lee2023comparison}. Although their teleport ranges differed from ours, their findings indicate a possible connection: larger thresholds for longer teleports could be related to decreased pointing accuracy for larger teleports. Both findings indicate that longer teleports are harder to fine-tune perceptually than small ones. However, teleport ranges are not classified cohesively across research, so comparison across studies is difficult  \cite{yao2023does}. This effect may also be influenced by distance estimation errors, which have been shown to increase with distance in VR \cite{gagnon2020far, ng2016depth}. Consequently, participants relying on distance estimation strategies may exhibit higher thresholds for longer teleports. However, this is dependant on the specific strategies participants use to detect whether their teleportation was adjusted or not.


We predicted that users with greater VR experience would exhibit smaller adjustment thresholds, as per \textbf{H3}, consistent with findings on sensitivity to redirected walking \cite{congdon2019sensitivity}. Our analysis revealed a significant negative correlation between VR experience and adjustment thresholds in the backward-small condition, indicating that more experienced users were better able to detect backwards adjustments for short-range teleports. A similar trend was observed for backward adjustments over the large ranges, though it did not reach significance. However, this effect may emerge fully with a larger sample size. Distance estimation in VR is generally biased, with distances typically underestimated \cite{el2019distance, renner2013perception}. It could be the case that users with more VR experience may be more accustomed to this distortion, enabling to more accurately detect adjustments. As forward adjustments generally had smaller perceptual thresholds, it appears that judging backward movements is more challenging. Our results indicate that VR experience is particularly advantageous for detecting backward adjustments. Therefore, H3, the hypothesis that users with greater VR experience would exhibit smaller adjustment thresholds, is supported in this context.


We included two measures of spatial ability in our study. The SBSOD provided a subjective assessment of participants’ spatial and navigational skills, while the SOT offered a more objective evaluation of their ability to imagine different perspectives. As per \textbf{H4}, we predicted that lower spatial ability would be associated with larger perceptual thresholds for teleport adjustments.

Our analysis revealed that only SOT scores were significantly correlated with perceptual thresholds, whereas no significant correlation was found for the SBSOD. This is not entirely unexpected as previous work has indicated weaker correlations of performance during teleportation tasks with the SBSOD compared to the SOT \cite{cherep2023individual}. Similar to the effects observed with VR experience, SOT correlations were apparent only for backward adjustment. Significance was found for backward teleports across the large range, though a similar but non-significant trend was observed for backwards adjustment across the small-range. The SBSOD did exhibit negative correlations across conditions, suggesting that a better sense of direction may be associated with lower perceptual thresholds. However as significance was not reached we cannot conclude this definitely. As only the SOT scores showed significant correlations, this could reflect the limitations of subjective measures (the SBSOD is measured from a self-reported questionnaire). Alternatively, the SOT may have tapped skills more relevant to the task. In particular, the SOT measures perspective-taking, which some participants reported using as a strategy when making teleportation decisions. This aligns with the demands of teleporting in VR, where users mentally simulate their viewpoint from a potential teleport destination. Nevertheless, further research is needed to determine whether the SBSOD can meaningfully predict sensitivity to adjusted teleportation, as such effects might emerge with a larger sample size.

Overall, backward adjustments exhibited larger perceptual thresholds, making them more difficult to detect. This may explain why higher perspective-taking ability and VR experience are associated with lower thresholds for these adjustments. In contrast, forward adjustments had smaller thresholds and appear to be universally detectable, regardless of participants’ experience or spatial ability. These findings highlight an interesting asymmetry in detectability between movement directions and warrant further investigation, ideally with a larger sample to confirm these trends.



Indeed while we examined teleportation adjustments in the context of perceptual thresholds, adjusting beyond thresholds could be considered. In the same line as personal space bubbles in VR, such adjusted teleports could be tolerated, or even welcomed as a game mechanic. Alternatively, as a locomotion method with a learning curve, adjusted teleports could be used to steer novice users. However, possible effects on user agency, comfort or presence should be considered.


This staircase method was our chosen psychophysical method in our experiment for a number of reasons. This method has been used in existing redirection research \cite{steinicke2009real, congdon2019sensitivity}. Additionally, since our task involved extensive locomotion in VR, we aimed to minimise participants’ time in the headset to reduce the risk of simulation sickness \cite{Duzmanska2018}. The staircase method was advantageous because of its efficiency, it concentrates trials around each participant’s threshold, reducing the total number of trials required \cite{cornsweet1962staircase}. This was suitable given we expected perceptual thresholds to be fine-tuned and individualistic. Given there was no research to form a basis for perceptual limits in teleportation, we did not have established bounds for a method of constant stimuli (MCS). However with the staircase, we could choose large bounds, and allow the bulk of the trials to converge at the individuals threshold.


\section{Limitation \& Future Work}

Our study revealed differences in perceptual thresholds for forward versus backward adjustments, as well as across more teleport ranges. As an initial study, we focused on forward and backward adjustments as these displacements were deemed most relevant for corrective teleport adjustments. The generalisability of these results to adjustments made in other directions (e.g., lateral movements) remains unclear. Future work should investigate such adjustments, which may exhibit lower perceptual thresholds due to more noticeable changes in the user’s viewing plane. Expanding the range of teleport distances, or exploring adjusted rotations would also be valuable. Different implementations of teleportation could also be considered for adjustments. Partially discordant teleportation (i.e. teleportation with physical rotation) has differing spatial consequences and could be examined for  influence on  perceptual thresholds \cite{cherep2020spatial, kelly2020teleporting}. Teleportation techniques that incorporate continuous visual transitions (e.g., dash teleportation) should be considered in future work. In such cases, not only the selected destination would be altered, but also the visual transition to that location. If users accept the adjusted destination, it is plausible that the corresponding transition would also be accepted, though this requires investigation. Importantly, these transitions introduce a continuous perceptual threshold, enabling better comparison to existing RDW perceptual thresholds.

The environment developed in this study enabled a controlled investigation of perceptual thresholds for adjusted teleports across varying displacement magnitudes. Translating these findings to social VR contexts requires consideration of additional factors.  In social VR, humanoid avatars could serve as anchoring cues, potentially influencing users’ perception of space and the detectability of teleport adjustments.Due to our experimental methods,  participants in our experiment were explicitly focused on teleportation. As such the reported thresholds likely represent conservative estimates. Thresholds may increase when attention is directed toward unrelated tasks \cite{nguyen2020effect, steinicke2009estimation}.

In social VR scenarios, users’ attention would be focused on their interaction partner rather than detecting adjustments. As a result, users may tolerate subtle redirection more readily, particularly since corrections for interpersonal space regulation are typically small (e.g., shifting a teleport from intimate to personal or social space \cite{hall1966hidden}). Our results indicate higher perceptual thresholds for backward adjustments, and prior work suggests that teleportation in social contexts is more prone to overshooting than undershooting \cite{connolly2025impact}. Together, these findings suggest that subtle backward teleport corrections can be applied with a low risk of detection in social VR settings. However, future work should examine how social context and engagement modulate these effects in applied VR settings.

Our methods also do not capture potential variability across environmental contexts. The room size was varied in response to the teleportation range and was not examined explicitly. Open-world environments, which lack boundary walls and provide fewer spatial constraints, may produce different perceptual thresholds. While we attempted to keep boundary and piloting cues consistent across teleports, variations in the shapes and sizes of objects may have unintentionally influenced participants’ ability to orient themselves.

The effects of such adjustments on users' navigational performance, and spatial orientation were not evaluated and should be investigated in future work. This is particularly relevant given our finding that participants with lower spatial orientation ability exhibited higher perceptual thresholds for backward adjustments, suggesting that individual differences may moderate the impact of adjusted teleportation. We also did not evaluate any such impact on user comfort. In social VR scenarios, subtly steering users away from uncomfortable interpersonal distances may positively influence comfort, warranting future investigation.

Finally, we did not systematically measure the potential effects of hardware or display factors, such as differences in FOV or resolution, which could influence participants’ perceptual sensitivity. Future studies should consider these factors, as variations in display characteristics might affect detection thresholds for adjustment.
\section{Conclusion}

This study presents Perceptually-Guided Adjusted Teleporting, showing how teleportation endpoints can be subtly altered without users noticing. Using a staircase method, we established baseline perceptual thresholds for forward and backward adjustments across small and large teleport ranges. Our results show that subtle, imperceptible adjustments are feasible, in particular for backwards adjustments, which may have applications in correcting overshooting. Perceptual thresholds were also found to increase with teleport range. Our results also suggest an individual nature to redirection sensitivity. Individuals with higher spatial abilities and experience in VR may be more sensitive to adjustments. Overall, our findings highlight adjusted teleportation as a promising approach for enhancing navigation and maintaining comfort in immersive VR environments.
\section*{Acknowledgments}
This work was conducted with the financial support of the Research Ireland Centre for Research Training in Digitally-Enhanced Reality (d-real) under Grant No. 18/CRT/6224 and RADICal (Grant No. 19/FFP/6409).


\bibliographystyle{abbrv-doi}

\bibliography{template}
\end{document}